\documentclass[sigconf]{acmart}
\usepackage{multirow}
\usepackage{subcaption}
\usepackage{graphicx}
\usepackage{hyperref}
\usepackage{array}
\usepackage{enumitem}

\usepackage{pgfplots}
\usepackage{tikz}
\usetikzlibrary{patterns}
\usepackage{calc}
\usetikzlibrary{calc}
\usetikzlibrary{shapes.geometric, arrows.meta, positioning}

\usepackage{booktabs}
\usepackage{pgfplotstable}
\usepackage{xcolor}
\pgfplotsset{compat=1.18}

\usepackage{listings}
\definecolor{stronglydisagree}{RGB}{64,64,64}      
\definecolor{disagree}{RGB}{128,128,128}           
\definecolor{neutral-neg}{RGB}{192,192,192}        
\definecolor{neutral}{RGB}{192,192,192}        
\definecolor{neutral-pos}{RGB}{192,192,192}        
\definecolor{agree}{RGB}{224,224,224}              
\definecolor{stronglyagree}{RGB}{255,255,255}      

\newcommand{\replicationurl}{https://tinyurl.com/2p8vre9u}

\makeatletter
\newcommand{\thickhline}{%
    \noalign {\ifnum 0=`}\fi \hrule height 1pt
    \futurelet \reserved@a \@xhline
}
\newcolumntype{"}{@{\hskip\tabcolsep\vrule width 1pt\hskip\tabcolsep}}
\makeatother

\AtBeginDocument{%
  \providecommand\BibTeX{{%
    \normalfont B\kern-0.5em{\scshape i\kern-0.25em b}\kern-0.8em\TeX}}}

\begin{document}


\title{Towards Personalizing Secure Programming Education with LLM-Injected Vulnerabilities}




 \author{Matthew Frazier}
 \affiliation{ 
 \institution{University of Delaware}  
\department{Computer and Information Sciences} 
 \city{Newark}
 \state{DE}
 \country{USA}
 }
 \email{matthew@udel.edu}

 \author{Kostadin Damevski}
 \affiliation{ 
 \institution{Virginia Commonwealth University}
 \department{Computer Science} 
 \city{Richmond}
 \state{VA}
 \country{USA}
 }
 \email{kdamevski@vcu.edu}



\begin{abstract}

According to constructivist theory, students learn software security more effectively when examples are grounded in their own code. Generic examples often fail to connect with students’ prior work, limiting engagement and understanding. Advances in LLMs are now making it possible to automatically generate personalized examples by embedding security vulnerabilities directly into student-authored code. This paper introduces a method that uses LLMs to inject instances of specific Common Weakness Enumerations (CWEs) into students' own assignment code, creating individualized instructional materials. We present an agentic AI framework, using autonomous LLM-based agents equipped with task-specific tools to orchestrate injection, evaluation, ranking, and learning outcome generation. 

We report the experience of deploying this system in two undergraduate computer science courses ($N = 71$), where students reviewed code samples containing LLM-injected vulnerabilities and completed a post-project survey. We compared responses with a baseline using a widely adopted set of generic security instructional materials. 
Students qualitatively reported finding CWE injections into their own code more relevant, clearer, and more engaging than the textbook-style examples. However, our quantitative findings revealed limited statistically significant differences, suggesting that while students valued the personalization, further studies and refinement of the approach are needed to establish stronger empirical support.

\end{abstract}

\ccsdesc[500]{Applied computing~Computer-managed instruction}
\ccsdesc[500]{Applied computing~Interactive learning environments}

\keywords{Computer Science Education, secure programming, software security, agentic workflows, CWE}



\maketitle

\vspace{-3mm}

\section{Introduction} 


A persistent challenge in secure programming education is that generic examples of software vulnerabilities often lack contextual relevance, making them easy to dismiss and difficult to internalize~\cite{wen_context_2019}. Students frequently struggle to connect abstract vulnerabilities to their own coding practices, limiting both engagement and comprehension~\cite{malinka2024using,fredricks2014eight}. This disconnect limits their ability to recognize, understand, and ultimately prevent real security flaws. To foster a more durable understanding, instructional materials must move beyond contrived textbook cases and instead situate vulnerabilities within contexts students find familiar and meaningful.

Constructivist learning theory holds that students build deeper understanding when new concepts are anchored in their prior knowledge and personal experience. This perspective is supported by research on learning engagement and competency-based education, which suggests that personally relevant content fosters stronger motivation and deeper learning~\cite{buckingham2012learning,cordova1996intrinsic,priniski2018making}. In the domain of software security, vulnerabilities embedded within student-authored code can create precisely this relevance, prompting students to recognize and reflect on their own design decisions. Yet realizing this personalization in an educational setting has historically been limited by instructor effort and available tooling.

The widespread availability of LLMs creates a powerful opportunity for software security education, where real-world vulnerabilities have traditionally been difficult to illustrate with clarity to students~\cite{yilmaz2022understanding,lam2022identifying}. Recent advances in agentic AI, the orchestration of autonomous language agents equipped with task-specific tools, enable personalized learning examples by using sophisticated, tool-driven reasoning pipelines over student-authored code. This emerging class of AI systems supports customizable workflows, offering a promising foundation for personalized instruction in computer science education.

To operationalize this potential, we introduce \textit{InjectEd}, 
a modular agentic AI system comprising of four agents: an Injector that embeds CWE-aligned vulnerabilities, an Evaluator that scores each injection’s pedagogical quality, a Ranker that selects the most suitable variant based on defined criteria, and a Learning Outcome Generator that produces learning assessment prompts. Each agent uses tools like syntax and semantic validators to inject vulnerabilities in ways that reflect the structure of student code.
InjectEd implements  a method for
injecting security vulnerabilities into student-authored code using LLMs configured to target specific Common Weakness Enumerations (CWEs). Our approach leverages model-generated edits that preserve local semantics while introducing representative security vulnerabilities into authentic project files. Each injection is tailored to a student’s codebase, producing a personalized artifact that connects the vulnerability to a student’s own design and implementation decisions. Ranking is grounded in four key dimensions (i.e., relevance, appropriateness, naturalness, and pedagogical value) ensuring that the targeted injection aligns with both the course context and instructional goals.

We explore the validity of this approach by deploying it in two upper-level university courses. We hypothesize that personalized CWE examples improve student understanding and perceived relevance, outperforming generic textbook samples in fostering engagement and secure programming skills. We report two main lessons from our deployment. First, it helped refine the educational design, which incorporates two key goals: G1) promoting personal relevance by connecting CWE manifestations to students’ own code, and G2) enhancing conceptual clarity through examples rooted in familiar development contexts. Second, the deployment allowed us to explore the experimental design and investigate the research question:  ``\textit{Do students better understand CWEs when shown examples from their own code compared to textbook-style examples?}''  Although we implemented all four agents with future classroom and research use in mind, the deployment focuses on the injector agent to evaluate the feasibility of using LLMs. 

The key contributions of this paper are:

\begin{itemize}[leftmargin=*, topsep=0.2em]
    \item novel method for using LLMs to inject personalized CWE-aligned vulnerabilities into student code;
    \item agentic AI system that leverages autonomous agents and tools to support scalable educational vulnerability injection;
    \item evaluation of student perceptions of relevance and conceptual clarity when exposed to personalized vs. generic examples.
\end{itemize}
\vspace{-3mm}

\section{Background and Related Work}

We summarize prior work on agentic AI architectures, LLM-based code injection, and challenges in secure programming instruction. Our work bridges these threads to explore LLM-driven personalization for security education.



\noindent
\textbf{Agentic AI and Educational Potential.}
Agentic AI refers to LLMs instantiated as autonomous agents with memory, role-based behavior, and tool invocation capabilities~\cite{wu_autogen_2023}. This design supports complex workflows through modular reasoning and multi-agent coordination, as seen in MetaGPT’s team simulations~\cite{hong_metagpt_2023}, AutoGen’s collaborative agents~\cite{wu_autogen_2023}, and CodeAgent’s automated security patching~\cite{tang_codeagent_2024}.

Despite its growth in developer tools, agentic AI remains largely absent in computing education. Prior systems rely on reactive prompting or static templates, offering hints but lacking contextual alignment with student-authored code. To our knowledge, no prior work in CS education deploys agentic AI for instructional intervention. By contrast, our approach leverages modular agents for tasks like vulnerability generation, relevance ranking, and learning outcome synthesis, grounded in example-based learning theory~\cite{wang2024large, taylor2016security}.

\noindent
\textbf{LLM-Based Vulnerability Injection.}
Recent work has explored LLMs for vulnerability injection, primarily to enhance static analysis training. Neural code transformers~\cite{nong_generating_2022}, edit-pattern mining~\cite{10172870}, and retrieval-augmented generation~\cite{daneshvar2024exploringragbasedvulnerabilityaugmentation} have improved the realism and coverage of seeded flaws. However, these techniques target automation rather than instruction. Educational systems rarely inject vulnerabilities to personalize learning. Our work introduces a novel use of LLMs to generate tailored, pedagogically grounded flaws in student-authored code.


\noindent
\textbf{Secure Programming Instruction.}
Despite curricular advances~\cite{bishop2011teaching, du2011seed, nestler2019nice}, secure programming remains conceptually difficult. Students often struggle with cognitive load~\cite{raina2014segmented}, lack of a security mindset~\cite{siraj2021there}, and the ability to identify or fix vulnerabilities~\cite{lam2022identifying, yilmaz2022understanding}. Instructional examples, critical for reducing these gaps, are often generic textbook snippets~\cite{taylor2016security} or uncontrolled bounty-style tasks~\cite{malinka2024using}, limiting personal relevance and clarity.

Studies show that personalized, context-rich feedback improves learning outcomes~\cite{wang2024large}. Although Copilot-style tools and agent-based tutors scaffold code synthesis and logic~\cite{whitney2018embedding}, few explore proactive injection of security flaws. Our work applies LLMs not just for correction, but for inserting pedagogical challenges directly into students’ own code to support deeper reasoning and transfer in security education.

\begin{figure*}[]
\centering
\begin{tikzpicture}[
    agent/.style={rectangle, rounded corners, draw=black, thick, fill=blue!15, minimum width=2.5cm, minimum height=1.0cm},
    artifact/.style={rectangle, draw=black, thick, fill=gray!10, minimum width=3cm, minimum height=1cm},
    infra/.style={rectangle, draw=black, dashed, thick, fill=yellow!15, minimum width=2.5cm, minimum height=0.cm},
    output/.style={rectangle, draw=black, thick, fill=green!10, minimum width=3.2cm, minimum height=0.9cm},
    arrow/.style={thick, -{Latex[length=3mm]}},
    node distance=0.8cm and 1cm
]

\node[artifact, align=center] (inputs) {Student Code +\\ Assignment Spec +\\ Course Desc. +\\ CWE List};

\node[agent, align=center, right=of inputs] (injector) {Injector\\Agent};
\node[output, align=center, above=of injector] (injectorout) {Injected Code +\\ CWE Metadata};

\node[agent, align=center, right=of injector] (evaluator) {Evaluator\\Agent};
\node[output, align=center, above=of evaluator] (evaluatorout) {Scored CWE\\Variants};

\node[agent, align=center, right=of evaluator] (ranker) {Ranker\\Agent};
\node[output, align=center, above=of ranker] (rankerout) {Best-Ranked\\CWE Injection};

\node[agent, align=center, right=of ranker] (generator) {Learning Outcome\\Generator Agent};
\node[output, align=center, above=of generator] (genout) {Formative Vulnerability\\ Questions};

\node[infra, align=center, below=of evaluator] (infra) {CrewAI + Langfuse + Similarity \& Semantic NLP Scoring Tools};

\draw[arrow] (inputs) -- (injector);
\draw[arrow] (injector) -- (evaluator);
\draw[arrow] (evaluator) -- (ranker);
\draw[arrow] (ranker) -- (generator);

\draw[arrow] (injector) -- (injectorout);
\draw[arrow] (evaluator) -- (evaluatorout);
\draw[arrow] (ranker) -- (rankerout);
\draw[arrow] (generator) -- (genout);

\draw[arrow] (infra.north) -- ++(0,0.3) -| (injector.south);
\draw[arrow] (infra.north) -- ++(0,0.3) -| (evaluator.south);
\draw[arrow] (infra.north) -- ++(0,0.3) -| (ranker.south);
\draw[arrow] (infra.north) -- ++(0,0.3) -| (generator.south);

\end{tikzpicture}
\caption{Overview of the InjectEd agentic pipeline. While only the Injector Agent was used in the current study, the full system includes Evaluator, Ranker, and Learning Outcome Generator agents, supported by shared infrastructure for prompt management and scoring.}
\label{fig:injected_architecture}
\end{figure*}
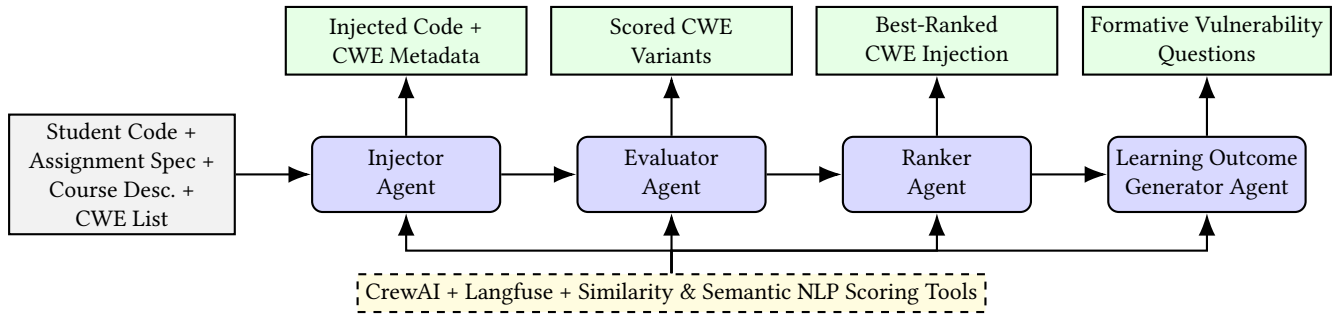

\section{Agentic CWE Injection System}


To automate personalized vulnerability injection for educational use, we designed the \textit{InjectEd} agentic AI system. Given a student’s recently written source code from an assignment submission, along with the corresponding course description and assignment specification, the system modifies the code to reflect a CWE-aligned vulnerability. 

InjectEd orchestrates four modular, sequential, and synchronous agents: the Injector, Evaluator, Ranker, and Learning Outcome Generator (see Figure~\ref{fig:injected_architecture} for an overview). Each agent performs a pedagogically motivated role using prompt-based reasoning, supported by shared infrastructure components such as CrewAI~\cite{crewAI} for orchestration, Langfuse~\cite{langfuse} for prompt management and observability, and internal tools for semantic and similarity scoring.

A pipeline controller coordinates data flow across agents, persisting outputs as JSON objects that conform to strict Pydantic schemas. These structured formats enable runtime validation, ensure compatibility with LLMs, and facilitate traceability. Agent prompts are versioned and retrieved dynamically from Langfuse, which also logs execution traces to support debugging and analysis. We use GPT-4o-mini across all agents to balance generation quality with cost efficiency. We provide a replication package of InjectEd at \replicationurl.

\subsection{Injector Agent}\label{sec:injector}

The Injector Agent introduces candidate CWE vulnerabilities into student-authored code while preserving syntactic and semantic integrity. Its primary objective is to produce pedagogically meaningful variations grounded in the original implementation context.

To identify viable modification regions, the agent first parses the student’s code using an abstract syntax tree (AST) analysis tool. This enables structural comprehension of the program and facilitates discovery of candidate injection points, such as unchecked input branches, unsafe function calls, or under-validated user interactions.

Given the known susceptibility of LLM generations for hallucination, especially when modifying executable code which has to be precise, the agent invokes a syntax validation tool post-injection to confirm that the output compiles or parses without error. This check suppresses hallucinated constructs and enforces structural correctness. A second tool compares behavioral outputs via a lightweight test suite to ensure that core functionality remains unchanged aside from the injected vulnerability. These validation steps reflect the broader design concern of maintaining authenticity and preserving pedagogical relevance without undermining program logic.

The agent outputs a structured JSON object that encapsulates the CWE identifier, descriptive metadata, modified source code, and a rendered HTML version suitable for classroom delivery. This output serves as the entry point for the downstream agent (i.e., Evaluator), who uses it to assess instructional quality and generate personalized feedback and assessment content. 


The agent ingests an exported CSV file of CWEs from the official CWE website where each row contains a CWE-ID, description, and metadata~\cite{CWE}, and iterates over all eligible entries during injection. This supports large-scale experimentation and avoids hardcoding a fixed CWE list.

\vspace{-3mm}

\subsection{Evaluator Agent}

The Evaluator Agent assesses each injected code variant for instructional quality across dimensions such as pedagogical relevance, cognitive load, and semantic coherence. It takes as input the original student code, the injected version, associated CWE metadata, and assignment context. Its goal is to determine whether the injection represents a realistic and useful teaching opportunity, rather than to assess correctness.

The agent uses structural and semantic comparison tools to support its judgment. It applies the Ratcliff/Obershelp algorithm to compute a normalized similarity score between the original and injected code~\cite{pan2025codellmsunderstanddesign}, capturing syntactic changes. A semantic drift score, based on Gensim embeddings~\cite{ishihara2022semantic}, reflects deeper meaning shifts beyond surface differences. These metrics appear in the prompt context, giving the language model concrete evidence to evaluate the impact of the injection on the student’s logic.

Evaluation criteria, grounded in CS education literature, include alignment with course objectives, naturalness of the flaw, and complexity relative to student skill. Prompting templates guide the model to generate 1–10 scores for each dimension, along with a free-text rationale. The agent also simulates likely misconceptions to assess whether the injection could provoke productive confusion, a key mechanism in example-driven learning.

This evaluative output drives downstream decisions. The Ranker Agent relies on these scores and rationales to select the most instructionally valuable injection. By quantifying pedagogical impact, the Evaluator ensures the system’s generative capacity remains aligned with sound instructional design.






\vspace{-3mm}

\subsection{Ranker Agent}
The Ranker Agent selects the most pedagogically valuable CWE injection from a pool of candidates, based on the Evaluator Agent’s multi-dimensional feedback. It ensures that only the most relevant and instructionally rich vulnerability is presented to the student. The agent receives injected code variants, evaluation scores, rationales, the original student code, assignment context, and CWE metadata.

Rather than relying solely on scores, the Ranker uses prompt-based comparative reasoning to synthesize quantitative and qualitative signals. Its prompts weigh tradeoffs, e.g., such as realism versus clarity or novelty versus assignment alignmen, and justify why a specific injection offers the highest instructional value. The agent uses internal tools including a scorecard-based ranking framework, simulated peer review prompts, and a student response generator to anticipate learner perception and difficulty.

These strategies address known LLM limitations like inconsistency and shallow evaluation. By combining structured heuristics with free-form rationale generation, the Ranker produces a more robust and explainable decision. Its output includes the top-ranked CWE injection and a narrative justification, which the Learning Outcome Generator uses to guide assessment design. The Ranker ensures the system delivers the injection most likely to support meaningful learning in context.







\subsection{Learning Outcome Generator Agent}

The Learning Outcome Generator is the final stage in the agentic pipeline, transforming the selected CWE-injected code into personalized formative assessments. Its goal is to scaffold conceptual reflection and reinforce secure programming principles through tailored learning artifacts. Using the injection and rationale from the Ranker Agent, it generates assessment items that contextualize the vulnerability within the student’s original code.

The agent produces multiple forms of instructional content: a short explanation of the CWE concept, a multiple-choice question (MCQ) with distractors based on common misconceptions, and an open-ended reflection prompt. Each item targets a range of cognitive skills, guided by Bloom’s taxonomy. Internal tools support this process by generating reflective prompts grounded in the code context, crafting MCQs with realistic distractors, and tagging each item with a Bloom level.

Outputs are structured for integration into course platforms or formative feedback loops, enabling instructors to deliver context-rich learning without manual effort. By closing the loop from injection to assessment, the agent ensures the system functions as a complete pedagogical intervention.

\begin{table*}[ht]
\caption{Post Survey Mean and St. Dev. of CWE Agent Injection and Baseline Preference (N=71). 
}
\begin{subtable}{0.5\textwidth}
    \centering
    \small
    \caption{Learning Impact}
    \begin{tabular}{||p{0.3\columnwidth}|p{0.14\columnwidth}|p{0.15\columnwidth}|p{0.17\columnwidth}||}
    \hline
    \textbf{Survey Question} & \textbf{Baseline} & \textbf{InjectEd} & \textbf{Significance} \\
     & mean (SD) & mean (SD) & (r, U-stat) \\
    \hline
Understanding & \textbf{4.4 (0.62)} & 4.27 (0.74) & ns, p > .05 \\ \hline
Fixing Capability & 3.73 (0.83) & \textbf{3.98 (0.76)} & ns, p > .05 \\ \hline
Confidence & \textbf{4.33 (0.71)} & 4.15 (0.85) & ns, p > .05 \\ \hline
Real-world Connection & 4.27 (0.69) & \textbf{4.41 (0.67) }& ns, p > .05 \\ \hline
Code Comprehension & \textbf{4.23 (0.73)} & 4.22 (0.69) & ns, p > .05 \\ \hline
Recognition Ability & \textbf{4.07 (0.94)} & 3.98 (0.79) & ns, p > .05 \\ \hline
    \end{tabular}
\end{subtable}%
\begin{subtable}{0.5\textwidth}
    \centering
    \small
    \caption{Perceived Engagement}
    \begin{tabular}{||p{0.23\columnwidth}|p{0.13\columnwidth}|p{0.15\columnwidth}|p{0.3\columnwidth}||}
    \hline
    \textbf{Survey Question} & \textbf{Baseline} & \textbf{InjectEd} & \textbf{Significance} \\
     & mean (SD) & mean (SD) & (r, U-stat) \\
    \hline
Relevance & 4.13 (0.68) & \textbf{4.22 (0.79)} & ns, p > .05 \\ \hline
Interest & 4.1 (0.76) & \textbf{4.37 (0.62)} & ns, p > .05 \\ \hline
Attention & 4.3 (0.7) & \textbf{4.37 (0.58)} & ns, p > .05 \\ \hline
Clear Illustration & \textbf{4.33 (0.8)} & 4.27 (0.59) & ns, p > .05 \\ \hline
Difficulty* & 2.53 (0.78) & \textbf{2.2 (1.05)} & ns, p > .05 \\ \hline
Confusion* & 2.37 (0.72) & \textbf{1.93 (1.03)} & p < .05 (r=2.53, U=832) \\ \hline
    \end{tabular}
\end{subtable}%
\label{tab:injected_significance}
\end{table*}

\section{Project Study}


Our study compares the educational impact of InjectEd’s \textit{CWE Injector Agent}, the first stage in our agentic system, with traditional textbook-style examples, assessing its effectiveness and pedagogical impact through mutliple metrics. While the full InjectEd architecture includes four agents (Injector, Evaluator, Ranker, and Learning Outcome Generator), this study focuses exclusively on assessing the pedagogical value of the Injector. To maintain high instructional integrity, the remaining agentic tasks were performed manually by the course instructor. Specifically, the instructor vetted the injected outputs for quality (acting as the Evaluator), selected the most pedagogically useful examples for each student (acting as the Ranker), and developed uniform reflection questions aligned with learning goals (substituting for the Learning Outcome Generator). This hybrid setup allowed us to rigorously evaluate personalized injection while preserving consistency and interpretability across participants. Student perspectives on their experiences are collected in post-project surveys.

\subsection{Methodology}

Our research question was: \textit{Do students better understand CWEs when shown examples in their own code versus textbook-style examples?} We hypothesized that CWE injection into a student’s own codebase would support greater conceptual clarity and personal relevance than abstract, textbook-style examples, based on constructivist learning principles~\cite{Rannikm}.


\subsubsection{Participants}

Participants were enrolled in one of two undergraduate computing courses at a U.S. university: Software Engineering and Database Systems. Each course involved semester-long final projects written in either Python, Java, Kotlin, or JavaScript. After final project submission, the course instructor distributed individualized links to a web-based InjectEd activity along with a post-project survey. Students were randomly assigned into one of two groups. Those in the treatment group viewed a version of their own project code modified by InjectEd to include a plausible CWE. Those in the control group received a matched, textbook-style example illustrating the same CWE, adapted from a publicly available pedagogical resource (Security Injections @ Towson) ~\cite{taylor2016security}.
~The InjectEd examples were delivered in HTML format for easy viewing and interaction. All participants received identical surveys regardless of group, and participation was voluntary and anonymous. This study was examined by our institution’s IRB and deemed exempt; as such, no formal consent forms were required.

\subsubsection{Process}

Prior to full deployment, the research team conducted several small-scale pilot runs to test InjectEd’s injection fidelity and HTML formatting. These early trials explored whether multiple CWE injections could be generated and then ranked to identify the most pedagogically useful examples. However, to minimize LLM bias and maintain conformance to course objective and coverage, the final study opted for a simplified design: the instructor curated a course-specific list of relevant CWEs, and three CWEs were randomly sampled for each student project. These were then injected into the two largest files in their final project (based on line and character count), resulting in six CWE-injected artifacts per project. The instructor reviewed the resulting injections and selected the most plausible and instructive file to present to each student.  Each CWE was deemed pedagogically relevant for the assignment, and randomization was used to avoid systematic exposure bias across students.

Following code delivery, students completed a validated post-project survey adapted from prior work.
The survey included twelve 5-point Likert-scale items measuring perceived understanding, clarity, relevance, and transferability of the example, as well as two open-ended questions prompting students to reflect on the usefulness of the example and whether they preferred personalized or textbook-style material.

\subsubsection{Data Analysis}

To evaluate differences in survey responses between the treatment and control groups, we used a Mann–Whitney U test, a non-parametric test appropriate for ordinal Likert data and small group sizes ($N = 71$). We computed the rank-biserial correlation coefficient ($r$) as a non-parametric effect size metric~\cite{kerby_2014_simple}. Survey items were grouped into two categories: \textit{Learning Impact} (e.g., understanding, confidence, fixing capability) and \textit{Perceived Engagement} (e.g., attention, relevance, confusion). Open-ended responses were thematically coded to identify recurring patterns related to conceptual clarity, personal relevance, and reflective insight. Given the small sample size, a formal inter-rater protocol was not feasible. Instead, the primary researcher coded all responses using grounded thematic analysis, a method consistent with prior CS education research.

\subsection{Results and Discussion}

Our results and discussion center around InjectEd's core design goals: (G1) increasing personal relevance by injecting CWEs into students' own projects; and (G2) improving conceptual clarity by illustrating security flaws within familiar code contexts.


\subsubsection{Quantitative Insights}

Table~\ref{tab:injected_significance} summarizes the post-survey results comparing students who received personalized InjectEd examples with those who received textbook-style baselines. 
While none of the learning outcome measures (i.e., perceived understanding, confidence, or ability to recognize or fix CWEs) achieved statistical significance ($p > .05$), students in the InjectEd condition consistently rated these dimensions slightly higher. In particular, responses trended favorably for ``Fixing Capability" and ``Real-world Connection," suggesting that personalized examples may facilitate deeper cognitive engagement, in line with goals G1 and G2.

The most notable quantitative result was a statistically significant reduction in reported confusion among InjectEd participants compared to the control group ($p < .05$, $r = 2.53$, $U = 832$), indicating a small to medium effect size. Although originally categorized as an engagement item, we interpret this outcome as evidence of improved conceptual clarity (G2), as lower confusion suggests that students were better able to comprehend the personalized example’s purpose and implications. Taken together, the directional improvements across learning measures and the significant decrease in confusion point to InjectEd’s potential to lower cognitive friction and support more accessible and context-sensitive learning, even if measurable gains in learning outcomes remain emergent.

Although most effects were not statistically significant, the directional trends across measures such as real-world connection and fixing capability suggest meaningful practical benefits. Given the modest sample size, lack of statistical significance should not be interpreted as lack of effect. Rather, the consistent student preferences and qualitative engagement themes indicate that InjectEd’s personalized examples may enhance cognitive access and instructional relevance in practice. A larger sample may be required to verify these trends quantitatively.

\subsubsection{Qualitative Insights.}

\input{figures/likert}

Open-ended responses reinforced InjectEd’s goals of personalization and conceptual clarity. Students frequently described the injection as ``eye-opening" or ``interesting," appreciating the unexpected introduction of realistic flaws into their own logic. These reactions suggest that the personalized CWE increased perceived authenticity and helped students reflect on overlooked security practices.

A recurring theme was enhanced engagement. Several students noted that the vulnerability made them ``think differently about their code" or realize they ``weren’t as secure as [they] thought." Even those already familiar with security principles acknowledged that seeing flaws introduced into their own work prompted deeper reflection than traditional examples.

Some students identified the injection’s instructional value in revealing specific vulnerabilities. One participant wrote that the flaw ``showed a type of bug I didn’t realize my logic was vulnerable to," highlighting the pedagogical strength of targeted, in-context examples. A few noted that the flaw felt ``subtle" or ``realistic," suggesting that the injection was both believable and impactful.

A minority of responses raised confusion, often tied to code complexity or perceived mismatch between the CWE and the assignment logic. These instances underscore the importance of contextual relevance for maximizing learning value. While not widespread, such feedback offers important design considerations for future injection selection strategies.

\subsubsection{Summary of What Worked Well and What Didn't Work Well}

InjectEd's injection agent performed well in practice and pedagogical impact. Its modular architecture enabled the generation of personalized CWE-injected examples in a structured HTML format, facilitating seamless integration into existing course workflows. The design supported high compatibility with instructor-led dissemination, and student feedback confirmed that the injected artifacts were both usable and impactful in the learning experience.

What worked particularly well was InjectEd's ability to embed vulnerabilities into student-authored codebases in a way that preserved contextual integrity. Students frequently described this integration as intuitive and relevant, with many highlighting the clarity gained from seeing mistakes within their own logic. This supports our design goals of fostering both conceptual clarity (G2) and personal relevance (G1). Additionally, the structured output allowed instructors to easily review and vet examples before distribution, effectively substituting for automated evaluation, ranking, and outcome generation in this early-stage deployment.

Grounding CWE selection in instructor-curated lists proved to be another strength. By constraining injections to CWEs directly aligned with course topics, the system avoided introducing irrelevant or overly complex flaws. However, to maintain fairness and reduce potential bias from the model, injected CWEs were selected at random from these approved lists. While this approach reduced algorithmic favoritism, it occasionally resulted in less pedagogically optimal matches. Some students received vulnerabilities that, while valid, felt less semantically connected to their submitted work. This mismatch was noted in open-ended responses and highlights the need for future refinement of injection selection, potentially through integration of a Ranking Agent or semantic matching system.

Despite generally positive feedback, a few implementation challenges emerged. Variability in injection quality was observed across some examples, which may be attributable to the inherent non-determinism of LLM-generated edits. Though instructor review helped filter problematic outputs, a small number of artifacts lacked clarity or felt forced. This suggests the need for an automated validation mechanism to standardize output and maintain a consistent instructional threshold.

Finally, while the InjectEd group consistently outperformed the control condition in terms of qualitative engagement and subjective learning experience, quantitative results did not yield widespread statistical significance. This is not uncommon in lightweight, one-time interventions. Future work could evaluate the system over multiple injections, projects, or instructional checkpoints to better capture its cumulative learning benefits. Expanding beyond a single injected artifact per student and incorporating reflective scaffolding may further surface InjectEd's full instructional potential.


\section{Threats To Validity} 

To support construct validity, our CWE injection tasks were informed by prior pedagogical frameworks~\cite{biggs2022teaching,anderson2001taxonomy,hindle2012naturalness}, and our survey instruments and injected examples were refined in collaboration with course instructors through a pre-evaluation process. Randomizing the CWE assignments and varying the injection/baseline ordering helped mitigate internal bias. While we used GPT-4o-mini for injection due to its strong general performance, results may differ with alternative or reasoning-augmented models. 
The primary external threat to validity lies in the limited generalizability of our findings, as the study included only 71 students across two undergraduate courses. Further replication across more diverse courses, institutions, and vulnerability types is recommended.

\section{Conclusions and Future Work} 

We introduced InjectEd, a system that uses agentic AI to inject personalized security vulnerabilities into student code, aiming to deepen understanding of Common Weakness Enumerations (CWEs). In a controlled study across two undergraduate courses, students reported increased relevance and reduced confusion. While most quantitative measures did not reach statistical significance, our qualitative findings suggest that personalized vulnerability injections can improve perceived relevance and reduce confusion.

Future work could include more comprehensive evaluation, improved alignment of injected CWEs with specific course objectives and assignment contexts, adaptive selection of vulnerabilities based on student code characteristics, and integration with instructor feedback mechanisms to enhance instructional effectiveness in secure software education.

\clearpage
\bibliographystyle{ACM-Reference-Format}
\bibliography{
    bib/citations,
    bib/seedvuln,
    bib/edu
}

@INPROCEEDINGS{hindle2012naturalness,
  author={Hindle, Abram and Barr, Earl T. and Su, Zhendong and Gabel, Mark and Devanbu, Premkumar},
  booktitle={2012 34th International Conference on Software Engineering (ICSE)}, 
  title={On the naturalness of software}, 
  year={2012},
  volume={},
  number={},
  pages={837-847},
  doi={10.1109/ICSE.2012.6227135}
}

@book{anderson2001taxonomy,
  title={A taxonomy for learning, teaching, and assessing: A revision of Bloom's taxonomy of educational objectives: complete edition},
  author={Anderson, Lorin W and Krathwohl, David R},
  year={2001},
  publisher={Addison Wesley Longman, Inc.}
}

@book{biggs2022teaching,
  title={Teaching for quality learning at university 5e},
  author={Biggs, John and Tang, Catherine and Kennedy, Gregor},
  year={2022},
  publisher={McGraw-hill education (UK)}
}

@article{kerby_2014_simple,
  title={The simple difference formula: An approach to teaching nonparametric correlation},
  author={Kerby, Dave S},
  journal={Comprehensive Psychology},
  volume={3},
  pages={11--IT},
  year={2014},
  publisher={SAGE Publications Sage CA: Los Angeles, CA}
}

@misc{pan2025codellmsunderstanddesign,
      title={Do Code LLMs Understand Design Patterns?}, 
      author={Zhenyu Pan and Xuefeng Song and Yunkun Wang and Rongyu Cao and Binhua Li and Yongbin Li and Han Liu},
      year={2025},
      eprint={2501.04835},
      archivePrefix={arXiv},
      primaryClass={cs.SE},
      url={https://arxiv.org/abs/2501.04835}, 
}

@inproceedings{ishihara2022semantic,
  title={Semantic shift stability: Efficient way to detect performance degradation of word embeddings and pre-trained language models},
  author={Ishihara, Shotaro and Takahashi, Hiromu and Shirai, Hono},
  booktitle={Proceedings of the 2nd Conference of the Asia-Pacific Chapter of the Association for Computational Linguistics and the 12th International Joint Conference on Natural Language Processing (Volume 1: Long Papers)},
  pages={205--216},
  year={2022}
}

@misc{crewAI, title={{CrewAI}: Framework for orchestrating role-playing, autonomous AI agents. by fostering collaborative intelligence, crewai empowers agents to work together seamlessly, tackling complex tasks.}, url={https://github.com/crewAIInc/crewAI}, journal={GitHub}, author={{CrewAI Inc}}, year={2025}}

@misc{langfuse, title={{Langfuse}: Open source LLM engineering platform: Traces, evals, prompt management, metrics, and playground to debug and improve your {LLM} application.}, url={https://github.com/langfuse/langfuse}, journal={GitHub}, author={{Langfuse GmbH / Finto Technologies Inc.}}, year={2025}}

@unknown{wu_autogen_2023,
author = {Wu, Qingyun and Bansal, Gagan and Zhang, Jieyu and Wu, Yiran and Zhang, Shaokun and Zhu, Erkang and Li, Beibin and Jiang, Li and Zhang, Xiaoyun and Wang, Chi},
year = {2023},
month = {08},
pages = {},
title = {AutoGen: Enabling Next-Gen LLM Applications via Multi-Agent Conversation Framework},
doi = {10.48550/arXiv.2308.08155}
}

@article{hong_metagpt_2023,
  title={Metagpt: Meta programming for multi-agent collaborative framework},
  author={Hong, Sirui and Zheng, Xiawu and Chen, Jonathan and Cheng, Yuheng and others},
  journal={arXiv preprint arXiv:2308.00352},
  volume={3},
  number={4},
  pages={6},
  year={2023}
}

@misc{tang_codeagent_2024,
      title={CodeAgent: Autonomous Communicative Agents for Code Review}, 
      author={Xunzhu Tang and Kisub Kim and Yewei Song and Cedric Lothritz and Bei Li and Saad Ezzini and Haoye Tian and Jacques Klein and Tegawende F. Bissyande},
      year={2024},
      eprint={2402.02172},
      archivePrefix={arXiv},
      primaryClass={cs.SE},
      url={https://arxiv.org/abs/2402.02172}, 
}

@misc{CWE,
  author={{The MITRE Corporation}},
  title={Common Weakness Enumeration: A community-developedlist fo {SW} \& {HW} weakness that can become vulnerabilities},
  url={https://cwe.mitre.org/},
  journal={CWE},
  year={2025},
  note={Retrieved on June 1, 2025}}

@misc{Rannikm, title={Social constructivism-jerome bruner}, url={https://link.springer.com/chapter/10.1007/978-3-030-43620-9_18}, journal={SpringerLink}, publisher={Springer International Publishing}, author={Rannikm{\"a}e, Miia and Holbrook, Jack and Soobard, Regina}, year={1970}, month={Jan}}

@incollection{malinka2024using,
  title={Using Real-world Bug Bounty Programs in Secure Coding Course: Experience Report},
  author={Malinka, Kamil and Firc, Anton and Loutock{\`y}, Pavel and Vostoupal, Jakub and Kristof{\'\i}k, Andrej and Kasl, Frantisek},
  booktitle={Proceedings of the 2024 on Innovation and Technology in Computer Science Education V. 1},
  pages={227--233},
  year={2024}
}

@article{wang2024large,
  title={Large language models for education: A survey and outlook},
  author={Wang, Shen and Xu, Tianlong and Li, Hang and Zhang, Chaoli and Liang, Joleen and Tang, Jiliang and Yu, Philip S and Wen, Qingsong},
  journal={arXiv preprint arXiv:2403.18105},
  year={2024}
}

@article{nestler2019nice,
  title={The NICE challenge project: providing workforce experience before the workforce},
  author={Nestler, Vincent and Coulson, Tony and Ashley, James D},
  journal={IEEE Security \& Privacy},
  volume={17},
  number={2},
  pages={73--78},
  year={2019},
  publisher={IEEE}
}

@article{taylor2016security,
  title={Security injections@{Towson}: Integrating secure coding into introductory computer science courses},
  author={Taylor, Blair and Kaza, Siddharth},
  journal={ACM Transactions on Computing Education (TOCE)},
  volume={16},
  number={4},
  pages={1--20},
  year={2016},
  publisher={ACM New York, NY, USA}
}

@article{bishop2011teaching,
	title={Teaching security stealthily},
	author={Bishop, Matt},
	journal={IEEE Security \& Privacy},
	volume={9},
	number={2},
	pages={69--71},
	year={2011},
	publisher={IEEE}
}

@article{du2011seed,
	title={{SEED}: hands-on lab exercises for computer security education},
	author={Du, Wenliang},
	journal={IEEE Security \& Privacy},
	volume={9},
	number={5},
	pages={70--73},
	year={2011},
	publisher={IEEE}
}

@inproceedings{raina2014segmented,
  title={Segmented and interactive modules for teaching secure coding: A pilot study},
  author={Raina, Sagar and Kaza, Siddharth and Taylor, Blair},
  booktitle={E-Learning, E-Education, and Online Training: First International Conference, eLEOT 2014, Bethesda, MD, USA, September 18-20, 2014, Revised Selected Papers 1},
  pages={147--154},
  year={2014},
  organization={Springer}
}

@article{whitney2018embedding,
  title={Embedding secure coding instruction into the ide: Complementing early and intermediate cs courses with eside},
  author={Whitney, Michael and Lipford, Heather Richter and Chu, Bill and Thomas, Tyler},
  journal={Journal of Educational Computing Research},
  volume={56},
  number={3},
  pages={415--438},
  year={2018},
  publisher={SAGE Publications Sage CA: Los Angeles, CA}
}

@inproceedings{siraj2021there,
  title={Is there a Security Mindset and Can it be Taught?},
  author={Siraj, Ambareen and Sridhar, Nigamanth and Hamilton Jr, John A Drew and Khan, Latifur and Kaza, Siddharth and Gupta, Maanak and Mittal, Sudip},
  booktitle={Proceedings of the Eleventh ACM Conference on Data and Application Security and Privacy},
  pages={335--336},
  year={2021}
}

@inproceedings{nong_generating_2022,
    author = {Nong, Yu and Ou, Yuzhe and Pradel, Michael and Chen, Feng and Cai, Haipeng},
    title = {Generating realistic vulnerabilities via neural code editing: an empirical study},
    year = {2022},
    isbn = {9781450394130},
    publisher = {Association for Computing Machinery},
    address = {New York, NY, USA},
    url = {https://doi.org/10.1145/3540250.3549128},
    doi = {10.1145/3540250.3549128},
    booktitle = {Proceedings of the 30th ACM Joint European Software Engineering Conference and Symposium on the Foundations of Software Engineering},
    pages = {1097–1109},
    numpages = {13},
    location = {Singapore, Singapore},
    series = {ESEC/FSE 2022}
}

@INPROCEEDINGS{10172870,
  author={Nong, Yu and Ou, Yuzhe and Pradel, Michael and Chen, Feng and Cai, Haipeng},
  booktitle={2023 IEEE/ACM 45th International Conference on Software Engineering (ICSE)}, 
  title={VULGEN: Realistic Vulnerability Generation Via Pattern Mining and Deep Learning}, 
  year={2023},
  volume={},
  number={},
  pages={2527-2539},
  doi={10.1109/ICSE48619.2023.00211}}

@misc{daneshvar2024exploringragbasedvulnerabilityaugmentation,
      title={Exploring RAG-based Vulnerability Augmentation with LLMs}, 
      author={Seyed Shayan Daneshvar and Yu Nong and Xu Yang and Shaowei Wang and Haipeng Cai},
      year={2024},
      eprint={2408.04125},
      archivePrefix={arXiv},
      primaryClass={cs.SE},
      url={https://arxiv.org/abs/2408.04125}, 
}

@article{yilmaz2022understanding,
	title={Understanding security vulnerabilities in student code: A case study in a non-security course},
	author={Yilmaz, Tolga and Ulusoy, {\"O}zg{\"u}r},
	journal={Journal of Systems and Software},
	volume={185},
	pages={111150},
	year={2022},
	publisher={Elsevier}
}

@inproceedings{lam2022identifying,
	title={Identifying Gaps in the Secure Programming Knowledge and Skills of Students},
	author={Lam, Jessica and Fang, Elias and Almansoori, Majed and Chatterjee, Rahul and Soosai Raj, Adalbert Gerald},
	booktitle={Proceedings of the 53rd ACM Technical Symposium on Computer Science Education V. 1},
	pages={703--709},
	year={2022}
}

@inproceedings{buckingham2012learning,
	title={Learning dispositions and transferable competencies: pedagogy, modelling and learning analytics},
	author={Buckingham Shum, Simon and Crick, Ruth Deakin},
	booktitle={Proceedings of the 2nd international conference on learning analytics and knowledge},
	pages={92--101},
	year={2012}
}

@article{cordova1996intrinsic,
      title={Intrinsic motivation and the process of learning: Beneficial effects of contextualization, personalization, and choice.},
      author={Cordova, Diana I and Lepper, Mark R},
      journal={Journal of educational psychology},
      volume={88},
      number={4},
      pages={715},
      year={1996},
      publisher={American Psychological Association}
}

@book{fredricks2014eight,
	title={Eight myths of student disengagement: Creating classrooms of deep learning},
	author={Fredricks, Jennifer A},
	year={2014},
	publisher={Corwin Press}
}

@article{priniski2018making,
  title={Making learning personally meaningful: A new framework for relevance research},
  author={Priniski, Stacy J and Hecht, Cameron A and Harackiewicz, Judith M},
  journal={The Journal of Experimental Education},
  volume={86},
  number={1},
  pages={11--29},
  year={2018},
  publisher={Taylor \& Francis}
}

@inproceedings{wen_context_2019,
    author = {Wen, Shao-Fang and Katt, Basel},
    title = {Learning Software Security in Context: An Evaluation in Open Source Software Development Environment},
    year = {2019},
    isbn = {9781450371643},
    publisher = {Association for Computing Machinery},
    address = {New York, NY, USA},
    url = {https://doi.org/10.1145/3339252.3340336},
    doi = {10.1145/3339252.3340336},
    booktitle = {Proceedings of the 14th International Conference on Availability, Reliability and Security},
    articleno = {58},
    numpages = {10},
    location = {Canterbury, CA, United Kingdom},
    series = {ARES '19}
}

\end{document}